
\magnification=1200
\def\pd{\partial}
\def\ts{\thinspace}
\rightline{DTP/93/31}
\rightline{May, 1993}
\vskip 20pt
\centerline{\bf  EQUATIONS OF HYDRODYNAMIC TYPE }
\vskip  20pt
\centerline{\bf{David B.Fairlie}}
\vskip 0.5 true cm
\centerline{Department of Mathematical Sciences}
\centerline{University of Durham, Durham, DH1 3LE, England}
\vskip 2 true cm
\noindent Abstract :

The universal field equations introduced by the author and his collaborators,
which admit infinitely many inequivalent Lagrangian formulations are
shown to arise as
consistency conditions for the existence of non-trivial solutions to
the quasi-linear equations, called equations of hydrodynamic type by
Novikov , Dubrovin and others. The solutions in closed form are only implicit.
A method due to Stokes, which is in essence
just Fourier Analysis is resurrected for application to those equations.
With the benefit of algebraic computation facilities, this method, allows the
general structure of power series solutions to be conjectured.

\vfill\eject

\centerline{\bf 1.Introduction.}
\smallskip
The purpose of this article is to comment upon some features of nonlinear
equations related to  the Euler equations of motion for fluid flow, which have
attracted considerable attention among theoretical physicists in recent months
after the seminal paper of Polyakov\rlap.$^{[1]}$. Two diverse lines of inquiry
regarding the properties of these equations are followed; The first is to
demonstrate how the equations which we have called Universal Field
Equations\rlap,$^{[2-5]}$ may be derived as consistency conditions on the
integrability of a simple set of first order equations which have been
extensively investigated by Dubrovin and Novikov\rlap.$^{[6,7]}$
and by Tsarev$^{[8-10]}$ under the name of `systems of hydrodynamic type'.
  The Universal Field Equations enjoy remarkable properties. They may be
derived from an infinite number of inequivalent Lagrangians, are associated
with an infinite number of conservation laws and are either reparametrization
invariant$^{[3]}$, or else the solutions are completely covariant. They may be
linearised by a Legendre Transform and thus integrated to give an implicit
solution$^{[5]}$.

The second  is concerned with an iterative method of solving such nonlinear
equations which goes back to Stokes$^{[11]}$, as referenced in Whitham's
book\rlap.$^{[12]}$. Stokes did not have the advantage of a sophisticated
algebraic manipulation package, such as MAPLE\rlap,$^{[13]}$ which permits the
rapid evaluation of approximations to a sufficient order to admit their
structure to be guessed, otherwise the development of methods of dealing with
integrable equations might have taken a different course. Consider first the
case of (1+1) dimensions; take the simple nonlinear wave equation,
$u_t =uu_x$, or perhaps better the KdV equation $u_t =6uu_x+u_{xxx}$, or the
Burgers equation, $u_t=uu_x+\nu u
_{xx}$, which is the Navier Stokes equation in the absence of a pressure
gradient in one space dimension. All of these equations can be solved in
principle; the first by the method of characteristics to give the implicit
solution $u=F(x+ut)$, the second by the Inverse Scatttering method and the
third by the Cole-Hopf transformation. In practice an explicit
approximate solution might be more useful, as frequently the problem posed is
that of the time evolution of an initial wave-form.
Look for a solution of the form of a formal Fourier expansion
$$u(x,t)=u_0(t)+ u_1(t)e^{ax}+u_2(t)e^{2ax}+u_3(t)e^{3ax}+\dots.\eqno(1.1)$$
The initial data is coded into the values of  the coefficients $u_j(t)$ at
$t=0$.
 Then the key point is that these coefficients are determined recursively  by a
set of inhomogeneous linear equations, where the driving term for the equation
for $u_n(t)$ is determined by terms of the coefficients $u_j(t),\  j<n$. $u_0$
is a constant and $u_1$ is determined
by a homogeneous linear equation; $u_2$ is then determined by an inhomogeneous
linear equation with driving term determined by $u_1$ and so on. The general
structure of the solution takes a remarkably simple form, and allows a Taylor
like expansion for solutions to the non-linear wave equation to be formally
deduced. If one looks for periodic solutions of the KdV equation, then this
method
gives a sequence of approximations to a given initial value problem. This is
just what the inverse scattering method was set up to achieve!  Admittedly this
construction is formal; but then the inverse scattering method permits a exact
solution in closed form to the initial value problem,  only when this
corresponds to a collection of of solitons! Indeed the multisoliton solution of
the KdV equation$^{[12]}$ can be recovered by formal summations using the above
 method\rlap.$^{[14]}$. Another possible application of this approach is to the
quantisation of such nonlinear equations, by considering the coefficients in
the expansion of (1.1) as creation operators.
\vskip 10pt
\centerline{\bf2. The Universal Field Equations as Consistency Conditions.}
\vskip 10pt
Consider the following set of equations
$$\sum_{j=1}^Nu_j{\pd u_i(x_k)\over\pd x_j}=0,\ \ i=1,\dots,N.\eqno(2.1)$$
A special case, in which $u(1)=-1,\ x_1=t$ constitutes
a multidimensional version of the the equations of hydrodynamic type$^{[7]}$
\rlap, or equivalently, (2.1) may be regarded as a static limit of  those
multidimensional equations. For the purposes of this section it is more
convenient to treat all the variables on the same footing, as in (2.1). Now
 suppose all $u_j,\ j=2,\dots,N$ are all functions of a
single function $u_1=v$. Then the result of differentiating (2.1) for $i=1$
 with respect to
$x_1,x_2,\dots x_N$ in turn  and using the easily proved result that
$${\pd u_i\over\pd x_j}{\pd v\over\pd x_k}={\pd u_i\over\pd x_k}{\pd v\over\pd
x_j}\eqno(2.2)$$
yields the $N$ equations
$$\eqalign{u_1v_{x_1x_1}+u_2v_{x_1x_2}+\dots+u_Nv_{x_1x_N}+&
({\pd u_1\over\pd x_1}+{\pd u_2\over\pd x_2}+\dots+{\pd u_N\over\pd
x_N})v_1=0\cr
u_1v_{x_1x_2}+u_2v_{x_2x_2}+\dots+u_Nv_{x_2x_N}+&
({\pd u_1\over\pd x_1}+{\pd u_2\over\pd x_2}+\dots+{\pd u_N\over\pd
x_N})v_2=0\cr
\dots\quad\dots\quad\dots+&\dots=0\cr
u_1v_{x_1x_N}+u_2v_{x_2x_N}+\dots+u_Nv_{x_Nx_N}+&
({\pd u_1\over\pd x_1}+{\pd u_2\over\pd x_2}+\dots+{\pd u_N\over\pd
x_N})v_N=0\cr}
\eqno(2.3)$$
Elimination of the $N+1$ variables $u_i,\  i=1,\dots N$ and
$\sum {\pd u_j\over\pd x_j}$ between the $N+1$ equations (2.1) and (2.3) yields
the determinantal condition
$$\det\pmatrix{0&{\pd v\over\pd x_k}\cr
               {\pd v\over\pd x_j}&{\pd^2 v\over\pd x_j\pd
x_k}\cr}=0\eqno(2.4)$$
This is just the Universal Field equation of reference [1]. Furthermore, the
extension to incorporate many fields arises in a similar manner. Suppose the
only the first $n$  among the $N$ functions  $u_j$ are functionally
independent;
i.e. the functions $u_j,\ j=n+1,\dots,N$ are functions of $u_i,\ i=1,\dots n.$
Then consider the equation
$$\sum_{j=0}^Nu_j\sum_{k=0}^n a_k{\pd u_k\over\pd x_j}=0.\eqno(2.5)$$
Here $\lambda_k,\ k=0,\dots,n$ are a set of arbitrary constants.
Differentiation with respect to $x_i$ gives
$$\sum_{j=0}^Nu_j\sum_{k=0}^n \lambda_k{\pd^2 u_k\over\pd x_i\pd x_j}+
\sum_{j=0}^n{\pd u_j\over\pd x_i}[\sum_k^n \lambda_k{\pd u_k\over\pd x_j}+
\sum_{r=n+1}^N{\pd u_r\over\pd u_j}\sum_{k=0}^n \lambda_k{\pd u_k\over\pd
x_r}]=0.\eqno(2.6)$$
Elimination of the $N+n$ variables $u_j$ and\hfill\break $(\sum_k^n
\lambda_k{\pd u_k\over\pd x_j}+
\sum_{r=n+1}^N{\pd u_r\over\pd u_j}\sum_{k=0}^n\lambda_k{\pd u_k\over\pd x_r})$
from the $N$ equations (2.6) {\it regarded as linear equations in those
variables} and the $n$ independent relations among the equations
$$\sum_{j=0}^Nu_j{\pd u_k\over\pd x_j}=0,\eqno(2.7)$$
 reproduces the
multifield form of the Universal equation. In compact form, these generalised
Bateman equations are given by
$$\det\pmatrix{0&{\pd u_a\over\pd x_j}\cr{\pd u_b\over\pd x_i} &\lambda^c{\pd^2
u_c\over\pd x_i\pd x_j}\cr} = 0\ .
\eqno(2.8)$$
Here, the determinant is that of a $(N+n\ts)\times (N+n\ts)$ matrix, with
$n\ts$ being the number of fields $u_a$ and $N\ts$ the number of
coordinates $x_i$ ($n < N\ts$), and the indices $(a,i)$ and $(b,j)$ refer
to rows and columns respectively (in particular, the entry ``$\ts0\ts$''
stands for the $n\times n$ null matrix). The quantities $\lambda^c$ are
arbitrary coefficients, in terms of which the determinant is to be expanded.
It is understood that (2.8) has to hold for {\it all} values of these
coefficients. Hence, one obtains $N-1\choose n-1$ equations -- clearly
generalising the Bateman equation (2.4) for one field to many -- whose general
covariance under arbitrary field redefinitions in $u^a$ is easily
established. Even though these equations form an over determined set, except
for $(n=1)$ or $(n=N-1)$ (the equations in the latter case were shown$^{[1]}$
to be also universal), their space of solutions is non empty$^{[1]}$, as any
arbitrary choice of functions $u_i(x_j)$, homogeneous of degree zero in the
variables $x_i$ will satisfy (2.8).
\vskip 10pt
\centerline{\bf 3. Introduction of explicit time.}
\smallskip

Consider the equations,
$$\eqalign {u_t + uu_x + vu_y = &0;\cr
v_t + uv_x + vv_y = &0.\cr}\eqno(3.1)$$

These equations are related to the one Polyakov  and others have studied in the
theory of conformal turbulence through setting $u = \psi_y; v = -\psi_x$. This
constraint ensures that the two dimensional velocity vector $\{u,v\}$ describes
incompressible flow; i.e. $u_x+u_y=0.$
 Of course the above equations simply say that the total derivative
${d\over dt} u(x,y,t) = 0$, and are the Euler equations for the free motion of
an inviscid fluid. The equation for the vorticity $\omega$, which is given by
$\omega = u_y-v_x=\nabla^2\psi$ which Polyakov studies is
$$\dot\omega=\psi_x\omega_y-\psi_y\omega_x\eqno(3.2)$$
There is a further consistency condition on $\psi$, arising from (3.1),
namely
$$\psi_{xx}\psi_{yy}-\psi_{xy}^2=0.\eqno(3.3)$$
if in addition to (3.2) equations (3.1) are imposed.
 These considerations have been extended to the Navier Stokes equation$^{[15]}$
by the intoduction of a viscous force in (3.2)
Somewhat curiously, if an irrotational flow with $\omega=0$ is sought, then
 $\{u,v\}$ is derivable from a potential $\phi$ as $\nabla\phi$, and the
equations (3.1) reduce to the following single equation for $\phi$,
$$\phi_t+{1\over2}(\nabla\phi)^2=0.\eqno(3.4)$$
Of course, the requirement of incompressibility imposes an additional
constraint upon $\phi$, namely that it should satisfy the equation
$\nabla^2\phi=0.$
It is easy to see, by transforming to complex variables $z=x+iy,\  \bar
z=x-iy$, that the only solution in this case is  $\{u,v\}$ = constant.
There are infinitely many Lagrangians;
$${f(uv)\over u}u_t+f(uv)u_x-f(uv)v_y. \eqno(3.5)$$
An implicit solution;
$$\eqalign{u=&F(x-ut,y-vt)\cr
           v =&G(x-ut,y-vt).\cr}\eqno(3.6)$$
where $F$ and $G$ are arbitrary functions of two variables can easily be
verified, and this result extends in an obvious manner to the solution of the
equations
$${\pd u_i\over\pd t} +\sum_{j=1}^Nu_j{\pd u_i\over\pd x_j}=0\eqno(3.7)$$
That these equations admit solutions which develop singularities can be seen by
looking for solutions which satisfy the ansatz
$$u_j(t,x_k) ={z_j(x_k)\over(t-a)}\eqno(3.8)$$
With the choice $z_i(x_k) = \sum A_{ik}x_k$, then (3.8) is a a solution
provided
the matrix $A$ satisfies $A=A^{\rm T}A$.

Introduce another set of fields $v_i,\ i=1\dots N$.Then
$${\cal L}=\sum_{j=1}^Nv_j{\pd u_j\over\pd t}+
\sum_{j=1}^N\sum_{k=1}^Nu_kv_j{\pd u_j\over\pd x_k}\eqno(3.9)$$
is a Lagrangian for the equations (2.4).
The companion set of equations, obtained by varying $\cal L$ with respect to
$u_i$ is
$${\pd v_i\over\pd t} +\sum_{j=1}^N(u_j{\pd v_i\over\pd x_j}+v_i{\pd
u_j\over\pd x_j}-v_j{\pd u_j\over\pd x_i})=0.\eqno(3.10)$$
\vskip 10pt
If $u$ is eliminated from  equations (3.1) the resulting equation for $v$ can
be written in a nice determinantal form:
$$\det\pmatrix{0&0&0&1&v\cr
0&0&v_x&v_y&v_t\cr
0&v_x&v_{xx}&v_{xy}&v_{xt}\cr
1&v_y&v_{xy}&v_{yy}&v_{yt}\cr
v&v_t&v_{xt}&v_{yt}&v_{tt}\cr}=0.\eqno(3.11)$$

 Note that if $v$ does not depend upon $y$ this equation is the
Bateman equation, various aspects of which are considered in references
[2-5].In that case the complete solution is given by the implicit solution for
$v$ of the equation
$$tf(v)+xg(v)=c\eqno(3.12)$$
where $f,\ g$ are arbitrary functions and $c$ is a constant which may be zero.
In the general situation the solution is given by the elimination of $u$ from
(3.6)
\vskip 10pt
\centerline{\bf 4. Explicit Formal Solution of the Nonlinear Wave Equation.}
\smallskip
It is well known that the general solution of the nonlinear wave equation
$$u_t =uu_x\eqno(4.1)$$
is solved implicitly as
$$u=G(x+ut).^{[12]}\eqno(4.2)$$
It may however be solved for an arbitrary differentiable initial value $F$
in terms of a Laplace, or Fourier expansion of the initial value of the form
$$F(x)=c_0+ c_1e^{ax}+c_2e^2{ax}+c_3e^{3ax}+\dots,\eqno(4.3)$$
by the method outlined in the introduction. $c_0$ may be taken as 1, by
rescaling $t$. A computer calculation using MAPLE$^{[13]}$ reveals the
conjecture that the power series expansion of $u(x,t)$ is given by
$$u(x,t)=\sum_{n_1,n_
2,\dots}\prod_{j=1}^{j=k} {c_j^{n_j}(Nt)^{n-1}\exp N(ax+t)\over
n_j!}\eqno(4.4)$$
Here $N=n_1+2n_2+\dots kn_k,\  n=n_1+n_2+\dots n_k.$
These coefficients may displayed as a formal contour integral, and the formal
sum constructed to give
$$\eqalign{
u(x,t)
=&1+\sum_{n=1}^{\infty}{1\over2\pi i}\oint
(c_1\exp a(x+z)+c_2\exp 2a(x+z)+\dots)^n({t\over (z-t)})^n{dz\over nt} \cr
=&F(0)+\sum_{n=1}^{\infty}t^{n-1}{\pd^{n-1}\over\pd
x^{n-1}}{(F(x+F(0)t)-F(0))^n\over n!},
\cr}\eqno(4.5)$$
after rescaling so that $F(O)=c_0$ is arbitrary.
The contour is taken sufficiently large so as to encircle the multiple poles at
$z=t$
The sum may itself be formally performed, to give the result
$$\eqalign{
u(x,t)=&1-{1\over2\pi i}\oint\log\Bigl(1-{t\sum_j c_j\exp aj(x+z) \over
(z-t)}\Bigr){dz\over t}\cr
=&F(0)-{1\over2\pi i}\oint\log(1-{t\over z-t}(F(x+F(0)z)-F(0)){dz\over
t}\cr}\eqno(4.6)$$
There does not appear to be a simple proof of the result (4.5), but it may be
used to obtain the explicit solution
$$u(x,t)={a+x\over1-t},\eqno(4.7)$$
which also follows fom (4.2).
It would be interesting to extend this to (3.1). We may note that if $v(t)$ is
any prescribed function of $u(t)$ the two
equations (3.1) reduce to a single equation,
$$u_t + uu_x + f(u)u_y = 0,\eqno(4.8)$$
where $f$ is arbitrary. This case at least, is not tractable by the above
method. However, the physically interesting case is when $u,\ v$ are derivable
from a stream function, and $u_x+v_y=0$. An explicit, possibly general,easily
verifiable class of solutions dependent upon an arbitrary function of a linear
combination of $x,\ y,\ t$ can be found by the Fourier method as
$$\eqalign{
u(x,y,t)=&u_0 +bG(ax +by+(au_0+bv_0)t),\cr
v(x,y,t)=&v_0 -aG(ax +by+(au_0+bv_0)t).\cr}\eqno(4.9)$$.
\vskip 10pt
\centerline{\bf 5. Conclusion}
Some properties of  the quasi linear system of equations of hydrodynamic type
have been explored, and their general solution given  in terms of arbitrary
functions, but only in implicit form. This is a feature of some nonlinear
completely integrable systems in higher dimension, and results from the fact
that one of the methods of solution is to interchange the role of dependent and
independent variables, whereupon the equation linearizes\rlap.$^{[5]}$
However, if a Fourier series solution, with time dependent coefficients is
sought, then
instead of the complete decoupling of the ordinary differential equations
describing the time evolution of the modes, these are determined successively
in terms of the fundamental mode. It is a remarkable feature that the structure
of the power series solution to the simplest nonlinear wave equation is easily
elucidated with the aid of an algebraic computation package, and a formal power
series solution obtained for an arbitrary initial configuration for the
simplest
nonlinear wave equation.
\smallskip

\vfill\eject
\centerline{REFERENCES}
\vskip 10pt
\frenchspacing
\item{[1]}  Polyakov  A.M. 1992  (The Theory of Turbulence in Two dimensions
{\it Princeton Preprint} (PUPT-1369)
\item{[2]}  Fairlie  D.B.,  Govaerts J. and  Morozov A. 1992 {\it  Nuclear
Physics}
\ {\bf  B373}\ 214-232
\item{[3]}
 Fairlie  D.B.  and Govaerts J. 1992 {\it Physics Letters}\ {\bf 281B}\ 49-53
\item{[4]}
 Fairlie  D.B.and Govaerts J. 1992 {\it J. Math Phys}{\bf 33}3543-3566
\item{[5]} Fairlie  D.B., J. Govaerts, 1993 Linearization of Universal Field
Equations, DTP-92/47 ( to appear in {\it J. Phys A})
\item{[6]} Dubrovin  B.A. and  Novikov S.P. 1983
{\it Soviet Math. Doklady}, {\bf 27} 665 ,1989\hfill\break
{\it Russ.Math.Surveys, } {\bf 44} 35
\item{[7]}   Dubrovin B.A. 1990 {\it Funct Anal and its Applications} {\bf 24}
280-285
\item{[8]}. Tsarev S.P. 1985 {\it Soviet Math. Doklady}, {\bf 31} 488
\item{[9]} Tsarev S.P. 1991 {\it Math. in the USSR, Izvestia} {\bf 37}  397-419
\item{[10]} Tsarev S.P. 1992 Classical Differential Geometry and Integrability
of Systems of Hydrodynamic Type, {\it Proceedings of NATO ARW `Applications of
Analytic and Geometric Methods to Nonlinear Differential Equations' 14-19 July
1992, Exeter, UK} hep-th preprint \# 9303092.
\item{[11]} Stokes G.G.1847 {\it Camb. Trans} {\bf 8}441-473
\item{[12]} Witham G.B.1974 {\it Linear and Nonlinear Waves}, Wiley
\item{[13]} Copyright 1992 Waterloo Maple Software
\item{[14]} Fairlie  D.B. and Mulvey J., work in progress.
\item{[15]} Migdal A.A.1993 Loop Equation in Turbulence  PUPT-1383
\end